# Functional Understanding Of Quantum Technology Is Essential To The Ethical Debate About Its Impact

Eline de Jong[1*]

## Abstract

As the innovative potential of quantum technologies comes into focus, so too does the urgent need of addressing their ethical implications. While many voices highlight the importance of ethical engagement, less attention has been paid to the conditions that make such engagement possible. In this article, I argue that *technological understanding* is a foundational capacity for meaningful ethical reflection on emerging technology like quantum technologies. Drawing on De Jong & De Haro's account of technological understanding (2025a; 2025b), I clarify what such understanding entails and how it enables ethical enquiry. I contend that ethical assessment, first and foremost, requires an understanding of what quantum technologies *can do*—their functional capacities and, by extension, their potential applications. Current efforts to build engagement capacities among broader audiences—within and beyond academic contexts—tend, however, to focus on explaining the underlying quantum mechanics. Instead, I advocate a shift from a *physics-first* to a *functions-first* approach: fostering an understanding of quantum technologies' capabilities as the basis for ethical reflection. Presenting technological understanding as an epistemic requirement for meaningful ethical engagement may appear to raise the bar for participation. However, by decoupling functional understanding from technical expertise, this condition becomes attainable for a broader group, contributing not only to a well-informed but also to a more inclusive ethical debate.

**Competing interests:** The author declares not having any conflicts of interest.

**Funding:** This publication is an outcome of the project Quantum Impact on Societal Security, project number NWA.1436.20.002, which is funded by the Dutch Research Council (NWO), The quantum/nanorevolution.

**Acknowledgements:** I am grateful to the organisers and participants of several workshops and conferences—EASST-4S (July 2024), the OZSW Annual Conference (August 2024), and *Understanding Science & Technology* (April 2024)—for the opportunity to present and discuss my work. I owe particular thanks to Dr Clare Shelley-Egan and Dr Sebastian De Haro for their thoughtful feedback on earlier drafts of this article, and to Professor Pieter Vermaas for insightful exchanges on the topic. These conversations have contributed significantly to the refinement of the paper.

---

[1] Institute for Logic, Language and Computation; Institute of Physics; Qusoft Research Center for Quantum Software; University of Amsterdam, The Netherlands

*Corresponding author: e.l.dejong@uva.nl

## 1. Introduction

Quantum technologies represent a new frontier in technological advancement, promising to offer capabilities that surpass many classical physics-based technologies. This potential, often termed "the quantum advantage", is associated with a host of opportunities and risks. Consequently, there have been numerous calls to scrutinise the ethical and broader societal aspects of these technologies, now while they can still be effectively shaped (Coenen & Grunwald, 2017; Inglesant et al., 2021; Hoofnagle & Garfinkel, 2022; Ten Holter et al., 2023; Seskir et al., 2023; Kop et al., 2024; Gasser, De Jong & Kop, 2024).

Although the calls for ethical engagement with quantum technologies have been loud and numerous, less attention has been given to the conditions that enable such engagement. Selin et al. (2017) argue that effective (public) engagement depends on a range of *capacities* —specifically, the knowledge and skills required for meaningful participation in discussions of science and technology. I argue that **technological understanding**—i.e. recognising how a technology can be used to achieve a practical aim—is one such enabling capacity: a foundational competence necessary for engaging in public discussions about technology. More specifically, I contend that technological understanding is a prerequisite for meaningful ethical debate. To study the impact and ethical implications of a technology, one must first understand the technology itself.

However, given quantum's reputation of being "too complex to understand", this may raise concerns about the possibilities for and accessibility of meaningful ethical discussion. In this article, I address these concerns by appealing to a multifaceted view of understanding. **To foster ethical discussion about quantum technologies, I advocate for promoting a specific kind of technological understanding—one focused on the capabilities of quantum technologies and their potential practical deployment, rather than their underlying physics.** First, I explain why an understanding of the technology at hand is crucial to scrutinising its ethical aspects. Second, I distinguish between three types of technological understanding, each typically applying to a different context. Third, I show that current efforts aimed at promoting understanding of quantum technologies tend to focus on the wrong type, that emphasises the underlying physics.

This article thus raises an epistemic threshold for ethical discussion of emerging technologies like quantum, emphasising the need to understand what a technology can do—but, by stressing the importance of understanding its functions rather than its underlying physics, it also makes this condition attainable for a wider audience, fostering an inclusive ethical debate.

## 2. Understanding as a critical enabler of ethical enquiry

Imagine a blue device, small enough to hold in your hand, with two buttons labelled "read" and "write". Would you feel sufficiently informed to discuss the ethical implications of this reader-writer? If I told you it was a "duplicator", you would still lack essential information—what exactly does it duplicate? Now, suppose I reveal that this device is designed to read electronic data from a car key and write it into a new key. These details are crucial to engage in a well-informed or simply meaningful discussion about ethical concerns raised by the duplicator.

A certain degree of knowledge about a technology seems indispensable for critical thinking about its ethical aspects and broader societal impact, without the need to be a technical expert. Yet, knowledge alone is not enough; it must be combined with the ability to analyse and interpret its implications. This aligns with Selin et al. (2017) who argue that effective public engagement with science and technology relies on (building) a range of "*capacities*". These capacities build on both knowledge and analytical skills. Drawing on these considerations, I propose that understanding—as the ability to *use knowledge*—is a key enabler of meaningful ethical engagement.

In the philosophical literature, understanding has typically been conceived as *the ability to do something with knowledge* (Reutlinger, Hangleiter & Hartmann, 2018; De Regt, 2017; Grimm, 2012; Wilkenfeld, 2013; Hills, 2016). Some have further specified it as the ability to make counterfactual inferences: to reason about "what-would-happen-if" scenarios (e.g., De Regt, 2017; Barman et al., 2024; De Jong & De Haro, 2025b). When mapping ethical issues regarding the potential impact of some (future) technology, such inferential abilities play a crucial role. In other words, ethical engagement can be considered a cognitive task, requiring the ability to use knowledge about the technology to construct "what-if" scenarios ("what if the technology materialises in a specific way?") and make "if-then" inferences ("if the technology has these capabilities, then those applications are possible").

The idea of understanding as an enabling skill for ethical discussion resonates with recent calls to build understanding of quantum technologies (Vermaas, 2017; Coenen & Grunwald, 2017: p.292; Hoofnagle & Garfinkel, 2022; p.xix; Seskir et al., 2023; Roberson, 2023; Rathenau, 2023: p.3; Quantum Delta Nederland, 2019: p.40). Vermaas (2017), for example, argues that understanding quantum technologies to "a reasonable degree" is a prerequisite to a societal debate about them. Yet such calls often leave unspecified what this discussion-enabling understanding actually involves and what kind of capacity we need to build. What type or level of understanding of quantum technologies is necessary to consider their ethical aspects?

## 3. Three types of technological understanding

To determine which type of understanding of quantum technologies is relevant for ethical discussions, it is helpful to first consider what it means to *understand* a technology. In response to this question, De Jong and De Haro (2025a; 2025b) introduced their notion of "technological understanding", defining it as *the ability or cognitive skill to recognise how a technology[2] can be used to realise a particular aim*. This ability requires that the technology in question is *intelligible* to the user: to successfully use a technology, one needs to the ability to prospectively reason about the consequences of operating it. This involves grasping the technology's properties and qualities, which can be understood at varying levels: technical, practical, or more conceptual.

Technological understanding thus requires some degree of insight into how the technology works and what it is capable of—though the required focus and depth of that insight will vary depending on the context in which the agent operates it and the specific goal pursued. In other words, what it means to have (enough) technological understanding depends on the context.

De Jong and De Haro identify three main contexts[3] that reflect and require technological understanding: the context of design, operation and innovation (2025a; 2025b). In each of these contexts, technological understanding is specified differently, resulting in three *types* of understanding with a specific focus in the required level of intelligibility of the technology and reasoning abilities.

In a design context, where the goal is to outline and build a technology, technological understanding involves reasoning about the artefact at the level of its inner workings. To design any technology—from a microwave oven to a quantum network—designers need an understanding of an artefact's principles and components in order to harness them. Such understanding is not necessarily required in the operation context, where the focus is on practical use. Here, technological understanding requires recognising the direct consequences of interaction. For instance, a driver must understand how pressing the pedals or using the gear stick affects the car's movement, without needing to comprehend the engine's mechanics. Similarly, operational understanding of a quantum sensing device involves knowing how to perform tasks with it, rather than how it functions internally.

In the context of innovation the central goal is to create a new relationship between a technology and a practical aim. While design involves "thinking up" the artefact itself, innovation, as defined by De Jong and De Haro, entails thinking up its instrumentalisation—determining how a technology can be applied. This includes both repurposing existing technologies—for example, leveraging the capabilities of graphic processing units (GPUs) to train artificial neural networks—and identifying new problems or needs that could be addressed by a yet-to-be-designed technology.

---

[2] De Jong and De Haro use the term 'technological artefact' to refer to a specific technology, contrasting it with technology in general. For readability, I choose to use '*a* technology' to refer to a particular technology.
[3] These contexts are not mutually exclusive and can, and often do, merge in practice.

The innovation type of technological understanding thus involves devising a technology's application: if researchers and engineers develop a large-scale quantum computer, what practical purposes could it serve? Answering such questions requires understanding the technology at the level of its functional capabilities and, by extension, assessing its fitness for purpose. In the next section, I argue that it is this functional understanding of (quantum) technology that is required for a meaningful ethical discussion about its implications.

Figure 1 summarises the three types of technological understanding:

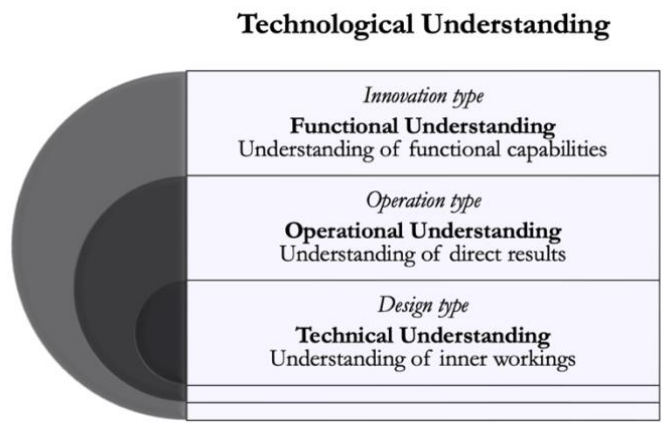

*Figure 1. Three types of technological understanding, each associated with a specific context and level of understanding.*

## 4. Understanding functions rather than physics

Thus far, I have argued that understanding is a key-enabling capacity for ethical engagement (Section 2) and proposed technological understanding as the relevant kind (Section 3). Building on this, I will now argue that ethical discussions about quantum technologies—particularly those focused on assessing impact—require *functional* understanding rather than an understanding of their underlying physics.

In practice, efforts aimed at fostering public understanding of quantum technologies tend to focus on cultivating an understanding of the underlying quantum mechanical phenomena. An empirical study by Van de Merbel et al. (2024) found that 81% of Dutch news articles about quantum technologies explain the physics behind quantum technologies. In the same study, the ability to recognise quantum phenomena is taken as an indicator of knowledge about quantum science and technology (Van de Merbel et al., 2024). Another study found that, in 70% of the TEDx talks given by experts about quantum science and technology, one or more quantum phenomena are explained (Meinsma et al., 2023). Apparently, introducing quantum technologies to a broader audience often involves a crash course on quantum mechanics.

This trend extends to academic and policy discussions: in this emerging field, many publications on the ethical and societal aspects of quantum technologies include sections on quantum phenomena such as superposition and entanglement (Hoofnagle & Garfinkel, 2022; Kop et al., 2024; Possati, 2023; Perrier, 2021; Rathenau Instituut, 2023; Vermaas, 2017; Top, 2024). These patterns reveal a 'physics-first approach' that prioritises scientific, technical explanations over a 'functions-first approach' that focuses on technological capabilities. This tendency is illustrated by Vermaas' specific call to philosophers of physics "for a renewed effort to make quantum theory understandable" in order to equip a broader group of stakeholders to engage in the societal debate on quantum technologies, suggesting that their expertise uniquely positions them to improve understanding among a wider audience (2017: p.242).

The ubiquity of explanations of the quantum mechanics behind quantum technologies reflects a broader tendency toward a technical, design-type understanding, which focuses on how physical phenomena interact with a technology's structure. While simplified explanations of quantum mechanics can spark public interest (Meinsma et al., 2024), they do not make quantum technologies epistemically accessible in a way that enables meaningful ethical discussion. After all, most people do not fully understand how combustion engines, transistors, or GPUs work, yet they can still engage in ethical debates about cars, smartphones, and AI. This suggests that technical understanding—focused on a technology's inner workings—is generally not essential for ethical reflection on its risks, benefits, and broader implications. Moreover, an overemphasis on physics, even in simplified form, may obscure rather than clarify quantum technologies, potentially hindering engagement and meaningful ethical discussions.

What is essential, however, is an understanding of quantum technologies' practical deployment—what they are capable of and how they might be used. To assess their potential impact, it is crucial to first envision the potential applications of quantum technologies. This ability aligns with functional understanding, which focuses on a technology's capabilities and, by extension, involves grasping the scope of its possible applications.

To foster meaningful ethical discussions about quantum technologies, especially in relation to their potential impacts, it is therefore vital to shift the focus from a physics-first to a functions-first understanding. Rather than centring on quantum mechanics, efforts to promote ethical engagement should focus on grasping how these technologies might function in practice. Only by achieving a clearer understanding of their capabilities can we enable truly meaningful discussions about their potential impact.

Prioritising functional understanding of quantum technologies for ethical engagement has two key implications. First, presenting understanding as a precondition for ethical discussions both raises and lowers the bar for participation: it establishes a necessary epistemic condition for engaging in such discussions, while simultaneously making this understanding more accessible by decoupling it from technical, physics-oriented expertise. Second, prioritising functional over technical understanding suggests that those with expertise in quantum mechanics, despite their deep knowledge, are not automatically well positioned to engage in ethical reflection. They too must acquire a distinct kind of understanding focused on the potential functions and applications of the technology.

## 5. Conclusion

The innovative potential of quantum technologies requires well-informed ethical discussions to ensure that its development and deployment is not only a technical success but also a net benefit for society and the planet. I have argued that effective ethical engagement is epistemically conditioned: it requires a certain degree of understanding of the technology. Technological understanding, therefore, should be seen as a precondition for meaningful ethical discourse.

This does not imply that ethicists or those involved in ethical discussions need to be technical experts—nor that technical experts are automatically well-positioned to engage in ethical reflection. What is crucial is having a grasp of the potential functionalities of quantum technologies to critically reflect on their ethical implications. Without such technological understanding, ethical discussions risk becoming overly speculative and detached from the realities of the technology. Therefore, fostering meaningful ethical discussions about quantum technologies requires shifting efforts from improving an understanding of their underlying physics to focusing on their potential functions. While this perspective raises an epistemic threshold for ethical discussions, it also makes them more accessible to a wider audience, fostering not only a well-informed but also a more inclusive ethical debate.